# Up-conversion injection in Rubrene/Perylene-diimide-heterostructure electroluminescent diodes

Ajay K. Pandey and Jean-Michel Nunzi[a]

*Equipe de Recherches Cellules Solaires Photovoltaïques Plastiques,*

*Laboratoire POMA, UMR-CNRS 6136,*

*Université d'Angers, 2 Bd Lavoisier, 49045 Angers, France*

[a] *Departments of Chemistry and Physics at Queen's University,*

*Kingston, Ontario, Canada,*

*nunzijm@queensu.ca*.

We implement and demonstrate in this paper a scheme that permits to drive electroluminescence with an extremely low turn-on voltage. The device behaves like compound semiconductors, in which the electroluminescence turn-on voltage is about the same as the open circuit voltage for the photovoltaic effect. However, the electroluminescence turn-on voltage is about half of the band gap of the emitting material, that cannot be explained using current models of charge injection in organic semiconductors. We hereby propose explanation through an Auger-type two-step injection mechanism.

Apart from their ease of fabrication and suitability to large area applications, there are many unexplored fields related to the peculiar physics of conjugated materials[1]. Of particular interest in that context is the recent discovery of the exceptionally low-threshold injection characteristics of rubrene crystals[2]. We implement here a resonant scheme that permits to drive electroluminescence (EL) with an extremely low turn-on voltage. The heterostructure device behaves like compound semiconductors, in which the EL turn-on voltage is about the same as the open circuit voltage for the photovoltaic effect. Surprisingly however, this places the EL turn-on voltage at about half of the rubrene band gap (2.2 eV). The physical interpretation can be found into the so-called Auger mechanism[3] that we could implement in our molecular heterojunction diode.

5,6,11,12-tetraphenylnaphthacene, commonly known as rubrene (Aldrich, sublimed grade, see insert in Fig. 1a), was used as a hole-transporting material, while N,N'-ditridecylperylene-3, 4, 9, 10-tetracarboxylic diimide (PTCDI, Aldrich, purum grade) was used as an electron transporting material. Both rubrene and PTCDI are widely studied semiconductors with among the highest field-effect mobility for holes[4] and electrons[5], respectively. Moreover, rubrene is also currently used as a yellow dopant for achieving efficient light emitting diodes (LED)[6] and efficient photovoltaic (PV) cells[7]. We have made heterojunction devices with 60 nm-thick rubrene and 20 nm-thick PTCDI layers sandwiched between transparent indium tin oxide (ITO, 15Ω/□, Merck) and 50 nm-thick aluminum electrodes. A 40 nm-thick poly-(3,4-ethylenedioxythiophene):poly-(styrenesulfonate) (PEDOT: PSS) at the ITO/rubrene interface and an 8 nm-thick bathocuproine (BCP) at the PTCDI/Metal interface were used as injection layers. For reference, rubrene-only devices were also made with 20 nm-Ca/40 nm-Ag metal cathodes. For comparison of the EL-characteristics also, PTCDI could be replaced by fullerene $C_{60}$ or $C_{70}$. All PEDOT-coated substrates were transferred inside an argon filled MBRAUN 200B glove box attached to an Edwards evaporation plant with

<0.1 ppm oxygen and moisture levels. All organic materials were grown at a constant deposition rate of 0.05nm/sec. All depositions were performed under a base pressure lower than $10^{-7}$ mbar. Device current-luminance-voltage (*J-L-V*) and current density-voltage (*J-V*) characteristics of the devices were recorded both in the dark and under white-light illumination using a Keithely 236 source measurement unit. Illumination was performed using an AM1.5 solar simulator (Steuernagel Solar constant 575). EL measurements were carried in ambient conditions on samples transferred without encapsulation at Thomson Multimedia Research Labs in Rennes. PEDOT: PSS and BCP layers that were used as passive buffer layers for stable and reproducible performance could be omitted from the device structure without affecting the results presented in the present paper.

Comparative current-luminance-voltage (*J-L-V*) characteristics of rubrene-only and rubrene/PTCDI heterojunction devices are shown in Figure 1. Rubrene only devices have an EL threshold at 2.2 eV (Figure 1a). They deliver a luminance of ~180 cd/m$^2$ at 2.4 V with (x = 0.55, y = 0.44) CIE color coordinates. Interestingly, rubrene/PTCDI heterojunction devices deliver visible yellow light at a driving voltage as low as 0.9 V. They emit with the same luminance and similar CIE coordinates at 1.4 V, what is ~1V below the single layer diode. The EL turn-on voltage of rubrene/PTCDI devices corresponds to half of the 2.2 eV rubrene band-gap. Forward current enters the mA-regime at drive voltages as low as 1 V (Figure 1b), providing enough carriers for recombination within the rubrene layer. Typical electroluminescence spectra of rubrene and rubrene/PTCDI diodes are shown in figure 2a. They are almost the same, as well as in rubrene/$C_{60}$ (lozenges in figure 2a) or $C_{70}$ devices (not shown). We also checked the current-voltage characteristics of rubrene/PTCDI diodes with 0.03 cm$^{-2}$-surface under 100mW/cm$^2$ AM 1.5 solar light. The open circuit voltage reaches $V_{OC}$ = 0.9 V under illumination. (Figure 2b).

Our bilayer devices behave like compound semiconductors, in which the electroluminescence turn-on voltage is about the same as the open circuit voltage for the photovoltaic effect. However, the electroluminescence turn-on voltage is about half of the band gap of the emitting material, that does not find explanation using current models of charge injection in organic semiconductors[8], as we discuss in the following. Under 1V bias, electron-hole pairs with ~1eV electrochemical potential recombining at the rubrene/$C_{60}$ interface must create ~2eV excitons in rubrene to yield effective electroluminescence with the characteristic color of rubrene (Fig.2a). To understand better that half-gap EL phenomenon, it is necessary to figure the energetic steps involved, starting from charge injection at the organic-metal interface. Figure 3a shows the energy level diagram of the present heterostructure. The energy level diagram of the rubrene-only device is also shown for comparison in figure 3b. The given HOMO-LUMO values are representative of the discrete rubrene[9] and PTCDI[10,11] layers. They may however shift significantly when used in contact with other organic layers[12,13]. As pictured in figure 3a, photo-induced charge transfer takes place under illumination from the rubrene electron donor the PTCDI electron acceptor. We get an open circuit voltage $V_{OC}$ = 0.9 V, which is approximately the rubrene/ PTCDI HOMO-LUMO offset in the flat band regime. In consequence, disorder at the heterojunction cannot be called to interpret the low EL-threshold voltage $V_{th}$ ($V_{th} \approx V_{OC}$). Thermal emission over the electron barrier from PTCDI-LUMO to rubrene-LUMO can be ruled out from the following equivalent arguments: First, it should take place more efficiently from the Ca electrode in contact with the single rubrene layer device in figure 3b than from the Al electrode in contact with PTCDI in figure 3a, that is not the case as figure 1 shows; Second, the energy barrier for thermal emission from PTCDI-LUMO to rubrene-LUMO is $F_i \approx$ 1eV and the thermal emission current density can be estimated from the Richardson-Dushman equation: $J_{th} = AT^2.\exp(-F_i /kT)$, where the theoretical value of $A$ is 123 $A.cm^{-2}K^{-2}$. One obtains $J_{th} \approx$ 50 $pA/cm^2$, very far from the $J \approx 1$ $mA/cm^2$ measured at $V$ = 1V (Figure 1b). EL at a relatively lower voltage than the semiconductor band-gap has been reported in distributed heterojunction polymer light emitting diodes[14]. It was identified as exciplex formation at the polymer-polymer interface, an energetically favorable situation for charge recombination without crossing the junction to the opposite polymer. As shown in Figure 2a, identical EL-spectral features for both single and bilayer devices rule out this possibility. In consequence, the half-gap voltage EL-threshold observed in Figure 1b bears all the features of an energy up-conversion process.

Energy up-conversion from charge injection requires the capture of energy by an electron located on an occupied level. A large energy can be produced by recombination of an electro-hole pair, a process commonly referred as Auger recombination. In semiconductor heterostructures and quantum dots, Auger recombination is the transfer of energy and momentum released by the recombination of an electron-hole pair to a third particle, ie. an electron (or hole), giving rise to an energetic electron (or hole)[3]. Auger recombination in inorganic semiconductors is obtained under the high carrier concentration regime provided by high doping or by high current density injection[15]. Traditionally, conjugated organic materials are low carrier concentration semiconductors and the occurrence of Auger recombination is less likely to occur in organics than in inorganic semiconductors. However, with the particular energy level conjunction that we get at the rubrene/PTCDI interface, where the PCTDI-LUMO is exactly midway between the rubrene HOMO and LUMO levels, a two-step excitation can take place, with the PTCDI-conduction band acting as a stepping stone.

Schematically, we propose the following mechanism: The rather large mobilities of the compounds and large energy-barrier at the heterojunction permits accumulation of charges of both signs at the interface under 1V-bias, which is almost the flat band regime (Fig.2b)[16]; Electrons from PTCDI recombine with holes from rubrene in a non-radiative process at the interface, with an excess energy of about 1 eV; That energy is transferred to an electron located on the PTCDI-LUMO (a PTCDI anion in other words[17]); it is just the energy that is required to resonantly excite that electron up to the rubrene-LUMO; The electron then recombines with the hole in the rubrene layer, emitting a photon.

Further improvement in the injection characteristics is anticipated from device optimization at various stages. In particular, improvement of the rubrene crystallinity should improve the current injection characteristics of the device by reduction of the trap density[2]. The use of dye-doped layers with high photoluminescence quantum yield may increase the EL efficiency[6]. Although we have described only rubrene/PTCDI heterostructures, our findings may easily be extended to other compositions bearing the same or similar resonance features. This efficient mechanism for charge injection and recombination into organic semiconductor structures at low drive (half-gap) voltages can be used as a building block for the realization of devices suitable for display, lighting and signalisation. Our findings also shed some light on the current understanding of organic semiconductor device physics.

**Acknowledgements**

We thank David Vaufrey and Henri Doyeux from Thomson R&D France, Research & Innovation, Rennes, for their help in EL characterization. We also thank Bernard Geffroy from CEA-LITEN in Saclay, for contradictory measurements. Financial support was obtained from the *Agence Nationale pour la Recherche* (ANR) under the solar photovoltaic program *NANORGYSOL*.

## Figure Captions

**Figure 1.** (a) Current (J) - Luminance (L) - Voltage (V) characteristics of *ITO /PEDOT /Rubrene /BCP /Ca* devices with EL turn-on at 2.2 V. Insert shows the chemical structure of Rubrene. (b) Current (J) - Voltage (V) characteristics of *ITO /PEDOT /Rubrene 60nm/PTCDi 20nm/BCP 8nm /Al 50nm* devices with EL turn-on at 0.9 V. Insert shows the same data in semi-log scale.

**Figure 2.** (a) Electroluminescence spectra near threshold of *rubrene only*, *rubrene/*$C_{60}$ and *rubrene/PTCDI* devices showing identical spectral features characteristic of rubrene emission with EL-peaks centered at 560 nm and 608 nm. (b) Photovoltaic response of *ITO /PEDOT /Rubrene /PTCDI /BCP /Al* device with $V_{OC}$ = 0.9 V.

**Figure 3.** (a) Energy level diagram for PV operation under open circuit showing relative HOMO-LUMO positions of rubrene and PTCDI layers with respect to PEDOT coated ITO and Al electrodes. (b) Energy level diagram for EL operation at threshold showing relative HOMO-LUMO positions of rubrene with respect to PEDOT coated ITO and Ca electrodes.

FIG.1.

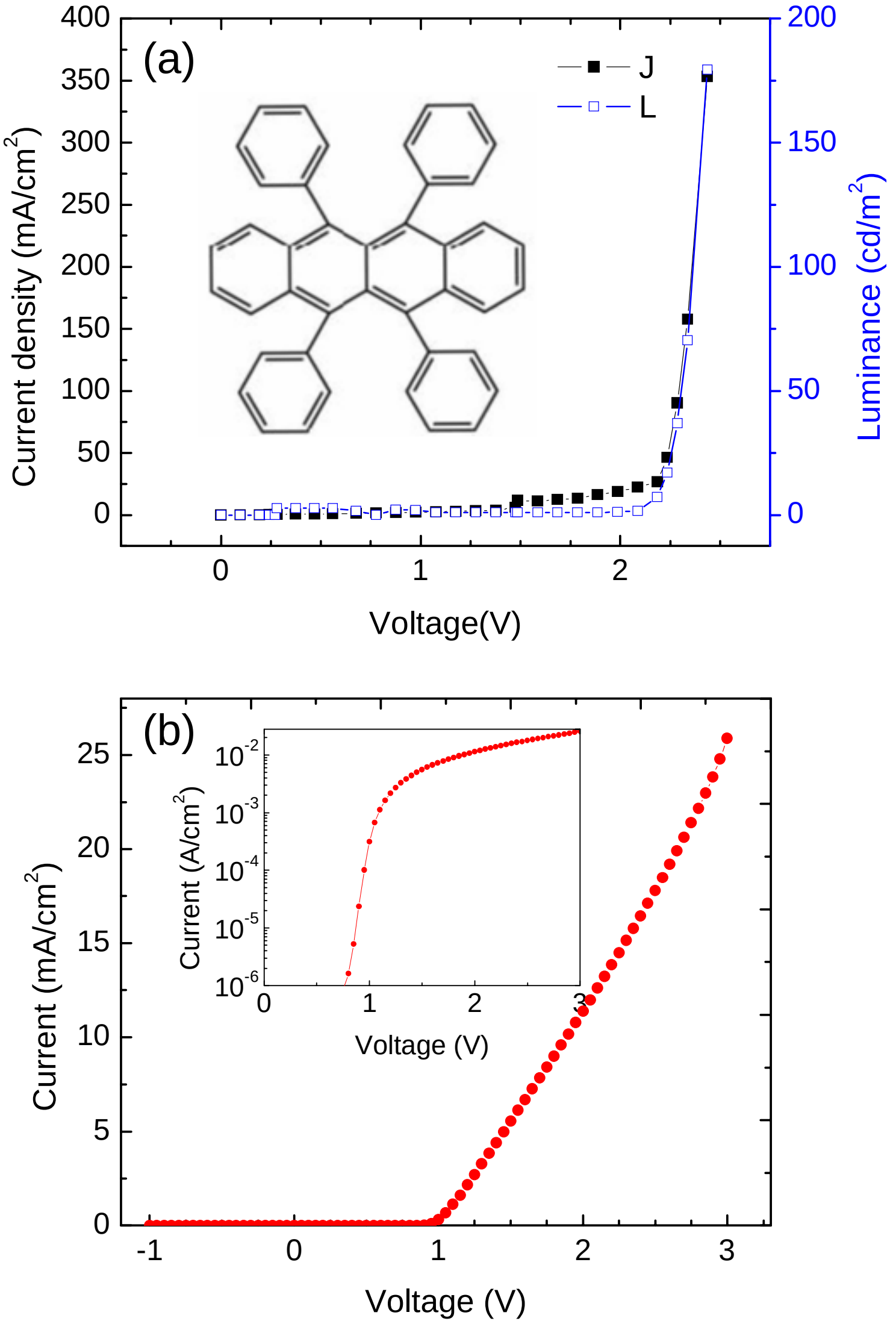

FIG.2.

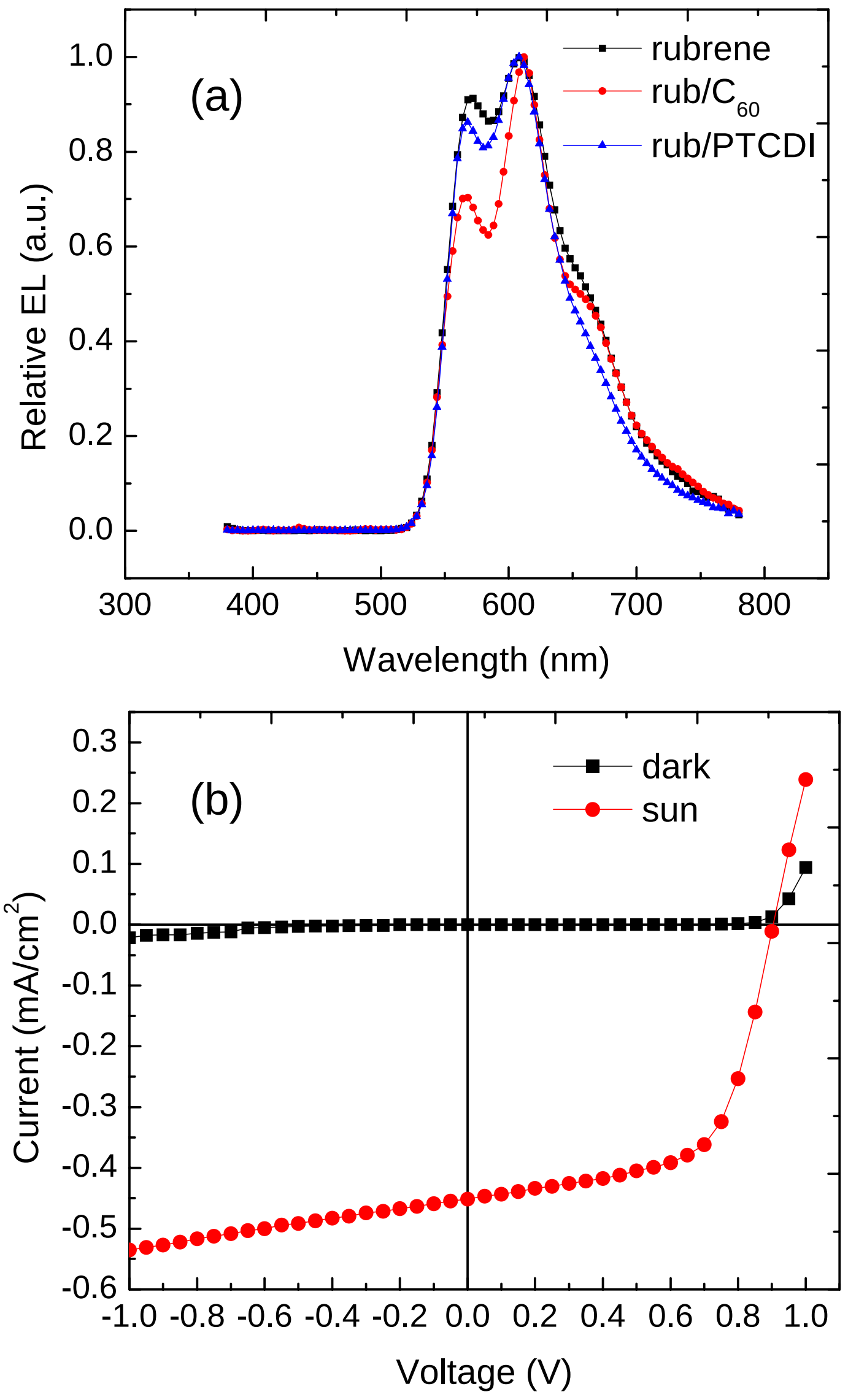

(a)
rubrene
rub/C60
rub/PTCDI
Relative EL (a.u.)
0.0
0.2
0.4
0.6
0.8
1.0
300
400
500
600
700
800
Wavelength (nm)
(b)
dark
sun
Current (mA/cm2)
0.3
0.2
0.1
0.0
-0.1
-0.2
-0.3
-0.4
-0.5
-0.6
-1.0 -0.8 -0.6 -0.4 -0.2 0.0 0.2 0.4 0.6 0.8 1.0
Voltage (V)

FIG.3.

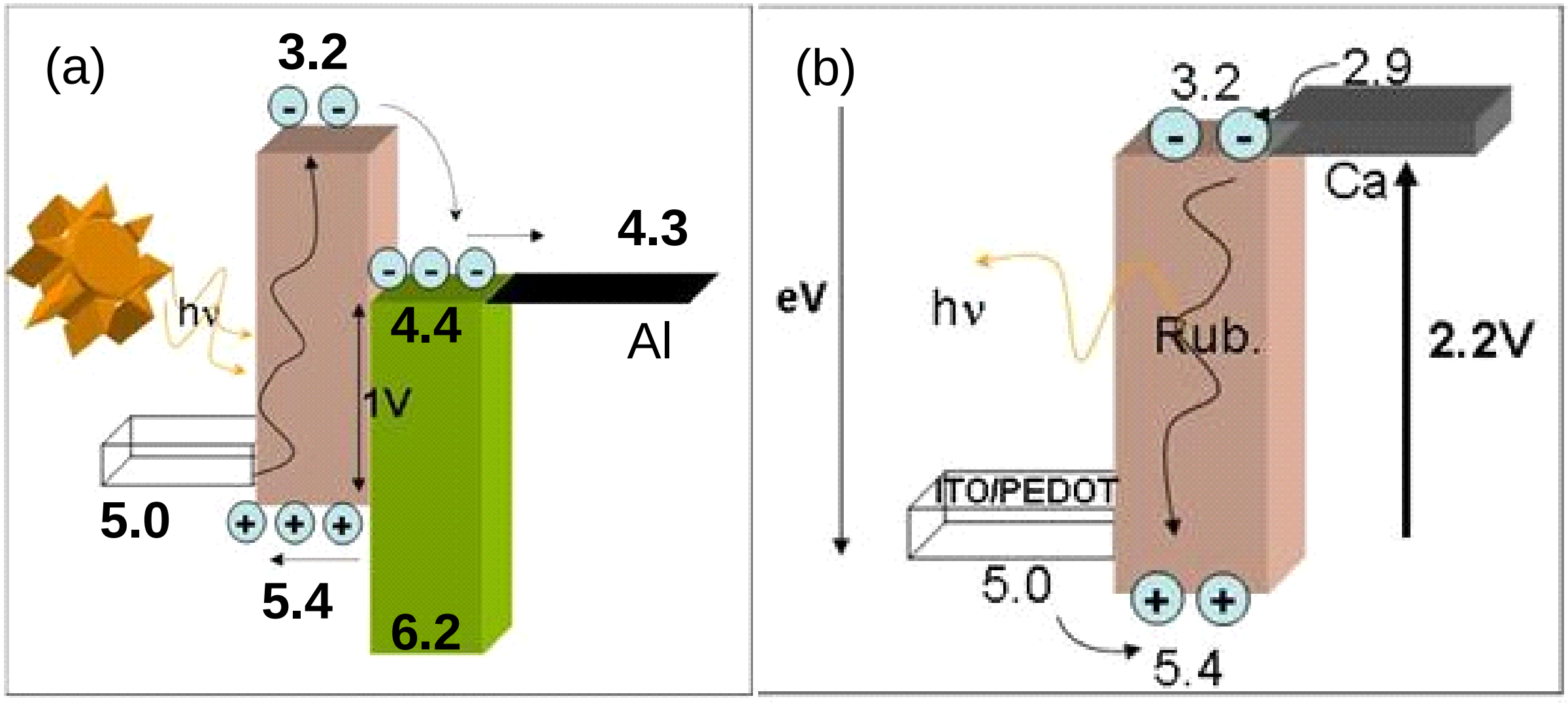